\documentclass[12pt]{article}
\usepackage{amssymb,graphicx,latexsym} 

\begin{document}

\newcommand{\sect}[1]{\setcounter{equation}{0}\section{#1}}
\renewcommand{\theequation}{\thesection.\arabic{equation}}
\newcommand{\be}{\begin{equation}}
\newcommand{\ee}{\end{equation}}
\newcommand{\bea}{\begin{eqnarray}}
\newcommand{\eea}{\end{eqnarray}}
\newcommand{\eps}{\epsilon}
\newcommand{\om}{\omega}
\newcommand{\vph}{\varphi}
\newcommand{\sig}{\sigma}
\newcommand{\CC}{\mbox{${\mathbb C}$}}
\newcommand{\RR}{\mbox{${\mathbb R}$}}
\newcommand{\QQ}{\mbox{${\mathbb Q}$}}
\newcommand{\ZZ}{\mbox{${\mathbb Z}$}}
\newcommand{\NN}{\mbox{${\mathbb N}$}}

\newcommand{\1}{\mbox{\hspace{.0em}1\hspace{-.24em}I}}
\newcommand{\II}{\mbox{${\mathbb I}$}}
\newcommand{\prt}{\partial}
\newcommand{\und}[1]{\underline{#1}}
\newcommand{\wh}[1]{\widehat{#1}}
\newcommand{\wt}[1]{\widetilde{#1}}
\newcommand{\mb}[1]{\ \mbox{\ #1\ }\ }
\newcommand{\half}{\frac{1}{2}}
\newcommand{\noin}{\not\!\in}
\newcommand{\rhotimes}{\mbox{\raisebox{-1.2ex}{$\stackrel{\displaystyle\otimes}
{\mbox{\scriptsize{$\rho$}}}$}}}
\newcommand{\bin}[2]{{\left( {#1 \atop #2} \right)}}
\newcommand{\ri}{{\rm i}}
\newcommand{\rd}{{\rm d}}
\newcommand{\A}{{\cal A}}
\newcommand{\B}{{\cal B}}
\newcommand{\C}{{\cal C}}
\newcommand{\F}{{\cal F}}
\newcommand{\E}{{\cal E}}
\newcommand{\cP}{{\cal P}}
\newcommand{\cN}{{\cal N}}
\newcommand{\R}{{\cal R}}
\newcommand{\T}{{\cal T}}
\newcommand{\I}{{\cal I}}
\newcommand{\V}{{\cal V }}
\newcommand{\W}{{\cal W}}
\newcommand{\cS}{{\cal S}}
\newcommand{\bS}{{\bf S}}
\newcommand{\cL}{{\cal L}}
\newcommand{\cV}{{\cal V}}
\newcommand{\hlp}{{\RR}_+}
\newcommand{\hlm}{{\RR}_-}
\newcommand{\Hil}{{\cal H}}
\newcommand{\D}{{\cal D}}
\newcommand{\G}{{\cal G}}
\newcommand{\U}{{\cal U}}
\newcommand{\form}{\langle \, \cdot \, , \, \cdot \, \rangle }
\newcommand{\e}{{\rm e}}
\newcommand{\by}{{\bf y}}
\newcommand{\bp}{{\bf p}}
\newcommand{\LL}{\mbox{${\mathbb L}$}}
\newcommand{\Rp}{{R^+_{\, \, \, \, }}}
\newcommand{\Rm}{{R^-_{\, \, \, \, }}}
\newcommand{\Rpm}{{R^\pm_{\, \, \, \, }}}
\newcommand{\Tp}{{T^+_{\, \, \, \, }}}
\newcommand{\Tm}{{T^-_{\, \, \, \, }}}
\newcommand{\Tpm}{{T^\pm_{\, \, \, \, }}}
\newcommand{\baral}{\bar{\alpha}}
\newcommand{\barbt}{\bar{\beta}}
\newcommand{\supp}{{\rm supp}\, }
\newcommand{\Pt}{\widetilde{P}}
\newcommand{\At}{\widetilde{A}}
\newcommand{\Bt}{\widetilde{B}}
\newcommand{\St}{\widetilde{S}}
\newcommand{\jt}{\widetilde{j}}
\newcommand{\Jt}{\widetilde{J}}
\newcommand{\Gt}{\widetilde{G}}
\newcommand{\xt}{\widetilde{x}}
\newcommand{\diag}{\rm diag}
\newcommand{\EE}{\mbox{${\mathbb E}$}}
\newcommand{\JJ}{\mbox{${\mathbb J}$}}
\newcommand{\MM}{\mbox{${\mathbb M}$}}
\newcommand{\ct}{{\cal T}}
\newcommand{\ph}{\varphi}
\newcommand{\phs}{\varphi^{(s)}}
\newcommand{\phb}{\varphi^{(b)}}
\newcommand{\phd}{\widetilde{\varphi}}
\newcommand{\phds}{\widetilde{\varphi}^{(s)}}
\newcommand{\phdb}{\widetilde{\varphi}^{(b)}}
\newcommand{\phl}{\varphi_{{}_L}}
\newcommand{\phr}{\varphi_{{}_R}}
\newcommand{\phpl}{\varphi_{{}_{+L}}}
\newcommand{\phpr}{\varphi_{{}_{+R}}}
\newcommand{\phml}{\varphi_{{}_{-L}}}
\newcommand{\phmr}{\varphi_{{}_{-R}}}
\newcommand{\phpml}{\varphi_{{}_{\pm L}}}
\newcommand{\phpmr}{\varphi_{{}_{\pm R}}}
\newcommand{\Ei}{\rm Ei}
\newcommand{\Fa}{\cal F(A)}
\newcommand{\Fb}{\cal F(B)}

\newcommand{\finprf}{\null \hfill {\rule{5pt}{5pt}}\\[2.1ex]\indent}

\pagestyle{empty}
\rightline{January 2010}

\vfill

\begin{center}
{\Large\bf Quantum Fields on Star Graphs\\ with Bound States at the Vertex}
\\[2.1em]

\bigskip

{\large
B. Bellazzini$^{a}$\footnote{bb424@cornell.edu}, 
M. Mintchev$^{b}$\footnote{mintchev@df.unipi.it} 
and P. Sorba$^{c}$\footnote{sorba@lapp.in2p3.fr}}\\

\null

\noindent 

{\it  
$^a$ Institute for High Energy Phenomenology
Newman Laboratory of Elementary Particle Physics,
Cornell University, Ithaca, NY 14853, USA\\[2.1ex]
$^b$ Istituto Nazionale di Fisica Nucleare and Dipartimento di Fisica, Universit\`a di
Pisa, Largo Pontecorvo 3, 56127 Pisa, Italy\\[2.1ex] 
$^c$ LAPTH \footnote{Laboratoire de Physique Th\'eorique d'Annecy-le-Vieux, 
UMR5108}, Universit\'e de Savoie, CNRS;\\ 
9, Chemin de Bellevue, BP 110, F-74941 Annecy-le-Vieux 
Cedex, France}
\vfill

\end{center}
\begin{abstract}

We investigate the propagation of a massless scalar field on a star graph, 
modeling the junction of $n$ quantum wires. The vertex of the graph is represented 
by a point-like impurity (defect), characterized by a one-body 
scattering matrix. The general case of off-critical scattering matrix 
with bound and/or antibound states is considered. We demonstrate that 
the contribution of these states to the scalar field is fixed by causality 
(local commutativity), which is the key point of our investigation. Two different 
regimes of the theory emerge at this stage. If bound sates are absent, 
the energy is conserved and the theory admits unitary time evolution. 
The behavior changes if bound states are present, because
each such state generates a kind of damped harmonic oscillator in the spectrum 
of the field. These oscillators lead to the breakdown of time translation invariance. 
We study in both regimes  the electromagnetic conductance of 
the Luttinger liquid on the quantum wire junction. We derive an 
explicit expression for the conductance in terms of the scattering matrix and 
show that antibound and bound states have a different impact, giving raise to 
oscillations with exponentially damped and growing amplitudes respectively.

\end{abstract}
\bigskip 
\medskip 
\bigskip 

\vfill
\newpage
\pagestyle{plain}
\setcounter{page}{1}

\sect{Introduction} 
\bigskip

Quantum field theory on graphs attracts recently much attention. 
Besides the purely theoretical interest, this attention 
is largely motivated by the physics of quantum wires. In fact, 
in many physical applications a quantum wire can be fairly well 
approximated by a graph. Each vertex of such a graph 
represents a junction of some wires. From the theoretical point of view 
the junction can be interpreted as a point-like
defect, which is characterized by a scattering matrix $S$. The goal of this 
paper is to explore the case when $S$ admits bound and/or antibound states. 
It is worth stressing that such states are absent 
at the scale-invariant (critical) points. In order to keep their  
contribution, one must explore therefore the theory away from criticality. 
In this respect the present investigation extends some previous studies \cite{NFLL}-\cite{Bellazzini:2009nk} 
in the subject, which have been mainly focused on the scale-invariant limit of the theory. 

We consider in this paper graphs $\Gamma$ of the type shown in Fig.1. They are 
called star graphs and represent the building blocks for generic graphs. We start by establishing 
the analytic structure of the scattering matrix $S$ associated with the vertex 
of $\Gamma$. In general $S$ admits both bound and antibound states. 
We demonstrate that their contribution to the massless scalar field $\ph$ is fixed by 
causality (local commutativity) up to a free parameter $t_m$. Two different regimes 
of the theory emerge at this stage. In absence of bound states, the field $\ph$ admits 
a unitary time evolution respecting time-translation invariance. 
The situation radically changes when $S$ admits bound states. Each of them generates in the spectrum of 
the theory a kind of damped harmonic oscillator. These oscillators lead to 
a breakdown of time-translation invariance. The energy of the system is no 
longer conserved, which signals a nontrivial flow of energy trough the boundary, being in this case 
the vertex of $\Gamma$. For $t<t_m$ the vacuum energy flow is outgoing and the system is in a 
dissipative regime. For $t>t_m$ the energy flow is incoming and the vacuum energy is growing. 
Although Hermitian, the underlying Hamiltonian is not self-adjoint, implying nonunitary time evolution. 
In what follows we refer to the bound states of $S$ as Boundary Bound States (BBS), 
because they decay exponentially in the bulk of the graph. We concentrate on the massless 
scalar field $\ph$ because it is the fundamental block of bosonization. 

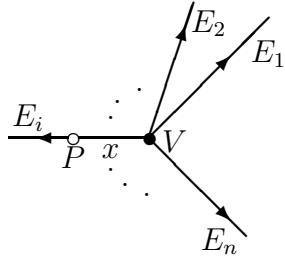
\begin{figure}[tb]
\setlength{\unitlength}{1mm}
\begin{picture}(20,20)(-42,20)
\put(25.2,0.7){\makebox(20,20)[t]{$\bullet$}}
\put(28.5,1){\makebox(20,20)[t]{$V$}}
\put(42,11){\makebox(18,22)[t]{$E_1$}}
\put(33,17){\makebox(20,20)[t]{$E_2$}}
\put(9,3.5){\makebox(20,21)[t]{$E_i$}}
\put(15,-1.2){\makebox(20,20)[t]{$P$}}
\put(20,-0.8){\makebox(20,20)[t]{$x$}}
\put(34.5,-12){\makebox(20,20)[t]{$E_n$}}
\thicklines 
\put(35,20){\line(1,1){16}}
\put(35,20){\line(-1,0){9.3}}
\put(24.4,20){\line(-1,0){8}}
\put(35,20){\line(1,-1){13}}
\put(35,20){\line(1,3){6}}
\put(20,3){\makebox(20,20)[t]{$.$}}
\put(20.9,5){\makebox(20,20)[t]{$.$}}
\put(23.8,6.6){\makebox(20,20)[t]{$.$}}
\put(20,-4){\makebox(20,20)[t]{$.$}}
\put(21.9,-6){\makebox(20,20)[t]{$.$}}
\put(24.8,-7.3){\makebox(20,20)[t]{$.$}}
\put(46,31){\vector(1,1){0}}
\put(46,9){\vector(1,-1){0}}
\put(40,35){\vector(1,3){0}}
\put(20,20){\vector(-1,0){0}}
\put(15,0.7) {\makebox(20,20)[t]{$\circ$}}
\end{picture} 
\vskip 1,5truecm
\caption{ A star graph $\Gamma$ with $n$ edges.} 
\label{stargraph}
\end{figure} 

In the second part of the paper we extend the bosonization procedure from the line 
to star graphs and analyze the impact of the off-critical boundary conditions on the Luttinger liquid. 
The renewed interest in Luttinger liquids arises from the recent progress in producing 
narrow quantum wires with few, or even single, conducting channels. Examples are 
quantum Hall edges \cite{Kang}, polydiacetylene \cite{Aleshin} and carbon 
nanotubes \cite{Terrones, Cumings}. As a potential application of our framework, 
we investigate the electromagnetic conductance of the Luttinger liquid on $\Gamma$ 
away from criticality. We derive an explicit expression of the conductance in terms of the 
scattering matrix and discuss in detail the different impact of bound and antibound states. 

The paper is organized as follows. In section 2 we describe the two regimes of a 
massless scalar $\ph$ field on $\Gamma$ with off-critical boundary conditions at 
the vertex. We discuss both the scattering and BBS contributions. 
Special attention is paid to local commutativity. We introduce here also the dual field $\phd$ and the vertex 
operators defined in terms of $\ph$ and $\phd$. In section 3 we derive the basic 
correlation functions. The vacuum instability due to the BBS and the accompanying 
breakdown of time-translation invariance are also analyzed there. 
In Section 4 we study the Luttinger liquid on $\Gamma$ and solve the 
Tomonaga-Luttinger model away from criticality. Coupling this model to a time-dependent 
electromagnetic field, we derive the conductance tensor in presence of (anti)bound states in $S$. 
We also show here the bound and antibound states imply different time behavior 
for the conductance, which represents a potential experimental signature for 
studying the analytic structure of the scattering matrix at the vertex of $\Gamma$. 
Section 5 collects our conclusions. In the appendices we give some computational details and 
prove some statements, used in the body of the paper.

\sect{Massless bosons with off-critical boundary conditions on a star graph} 

\subsection{Preliminaries}

We consider in this section the equation 
\be
\left (\prt_t^2 - \prt_x^2 \right )\ph (t,x,i) = 0\, , 
\qquad x> 0 \, ,  \;  i=1,...,n \, . 
\label{eqm1}
\ee 
where $t$ is the time and $(x>0,i=1,...,n)$ are the coordinates of any point $P$ 
on the star graph $\Gamma$, shown in Fig.~\ref{stargraph}. 
In order to solve (\ref{eqm1}), we must fix the 
boundary condition in the vertex (junction) $V$ of $\Gamma$. 
A standard QFT assumption at this point is to require that the operator 
$K\equiv -\prt_x^2$ on $\Gamma$ is 
self-adjoint\footnote{In order to keep the BBS contribution we do not 
impose $K\geq 0$.}. This requirement, which is trivially satisfied on the line $\RR$, 
is delicate on $\Gamma$. Fortunately however, the problem has been 
intensively investigated in the recent mathematical literature
\cite{Kostrykin:1998gz}-\cite{KSch}, where the subject goes under 
the name of ``quantum graphs" \cite{EKKST}. From 
these studies one infers that $K$ is self-adjoint on $\Gamma$ 
if and only if the field $\ph$ satisfies the 
boundary condition 
\be 
\sum_{j=1}^n \left [\lambda (\II-U)_{ij}\, \ph (t,0,j) -\ri (\II+U)_{ij}
(\prt_x\ph ) (t,0,j)\right ] = 0\, , 
\qquad \forall \, t\in \RR\, , \; i=1,...,n\, , 
\label{bc} 
\ee 
where $U$ is any unitary matrix and $\lambda > 0$ is a 
parameter with dimension of mass introduced in order to 
recover the correct physical dimensions. 

Eq. (\ref{bc}) generalizes to the graph $\Gamma$ the  
mixed (Robin) boundary condition on the half-line $\RR_+$. 
The matrices $U=\II$ and $U=-\II$ define the Neumann and Dirichlet 
boundary conditions respectively. The physical interpretation of 
(\ref{bc}) in the context of bosonization is established in \cite{Bellazzini:2008mn}. 
At criticality (\ref{bc}) describes the splitting of the $U(1)$-charge current at the junction $x=0$. 
We show below (see eqs. (\ref{lbc}), (\ref{lbc1})) that (\ref{bc}) can be reformulated in terms 
of currents also away from criticality.  

Besides the boundary conditions, we must fix also the initial conditions imposing 
the equal-time commutation relations 
\be
[\ph (0,x_1,i_1)\, ,\, \ph (0,x_2,i_2)] = 
[(\prt_t\ph )(0,x_1,i_1)\, ,\, (\prt_t\ph ) (0,x_2,i_2)] = 0\, , \qquad 
\label{initial1}
\ee
\be 
[(\prt_t\ph )(0,x_1,i_1)\, ,\, \ph (0,x_2,i_2)] = 
-\ri \delta_{i_1i_2}\, \delta (x_1-x_2) \, .
\label{initial2}
\ee 

As already mentioned in the introduction, for the explicit construction 
of $\ph$ it is convenient to interpret \cite{Bellazzini:2006jb} the vertex of $\Gamma$ as a
point-like impurity (defect) \cite{Delfino:1994nr}-\cite{Mintchev:2003ue}, 
characterized by a nontrivial scattering matrix $S$. 
The $S$-matrix is associated \cite{Kostrykin:1998gz} to the operator $K$ on $\Gamma$
and is fully determined by the boundary conditions (\ref{bc}). The 
explicit form of $S$ is \cite{Kostrykin:1998gz, H1}
\be 
S (k) = -[\lambda (\II - U) + k(\II+U )]^{-1} 
[\lambda (\II - U) - k(\II+U )] 
\label{S1}
\ee 
and has a transparent physical meaning: the diagonal element $S_{ii}(k)$ 
represents the reflection amplitude from the defect on the edge $E_i$, whereas  
$S_{ij}(k)$ with $i\not=j$ equals the transmission amplitude from $E_i$ to $E_j$. 

By construction (\ref{S1}) is unitary 
\be 
S(k)^*=S(k)^{-1} \, , 
\label{unit1}
\ee 
and satisfies Hermitian analyticity 
\be 
S(k)^*=S(-k)\, .  
\label{Ha}
\ee 
Combining (\ref{unit1}) and (\ref{Ha}), one gets 
\be 
S(k)\, S(-k) = \II  \, . 
\label{unit2} 
\ee 
Notice also that 
\be 
S(\lambda ) = U \, . 
\label{S2}
\ee 
We see that the boundary condition (\ref{bc}) is fixed actually by the scale $\lambda$ and the value of 
scattering matrix at that scale. 

Let us establish now the analyticity properties of (\ref{S1}) in the 
complex $k$-plane. Let $\U$ be the unitary matrix 
diagonalizing $U$ and let us parametrize 
\be 
U_d=\U^{-1}\, U\, \U 
\label{d1}
\ee  
as follows 
\be 
U_d = \diag \left (\e^{2\ri \alpha_1}, \e^{2\ri \alpha_2}, ... , \e^{2\ri
\alpha_n}\right )\, , \qquad \alpha_i \in \RR\, . 
\label{d2}
\ee 
Using (\ref{S1}), one easily verifies that $\U$ 
diagonalizes also $S(k)$ {\it for any} $k$ and that 
\be 
S_d(k) = \U^{-1} S(k) \U = 
\diag \left (\frac{k+\ri \eta_1}{k-\ri \eta_1}, \frac{k+\ri \eta_2}{k-\ri \eta_2}, ... , \frac{k+\ri \eta_n}{k-\ri \eta_n} \right )\, , 
\label{d3}
\ee 
where 
\be 
\eta_i = \lambda \tan (\alpha_i)\, , 
\qquad -\frac{\pi}{2} \leq \alpha_i \leq \frac{\pi}{2}\, .  
\label{p1}
\ee 
We have therefore a simple direct proof of the following statement \cite{KSch}: 

{\bf Proposition 1:} {\it The scattering matrix $S(k)$ given by (\ref{S1}) is a meromorphic function 
with poles located on the imaginary axis and different from 0.} 

The poles belonging to the lower and upper half-plane are known \cite{RS} as 
{\it antibound} and {\it bound} states respectively. 
It is natural to represent at this stage the field $\ph$ as a linear combination 
\be 
\ph (t,x,i) = \phs (t,x,i) + \phb (t,x,i) \, , 
\label{dec1}
\ee 
where $\phs$ collects the contribution of the {\it scattering} states and 
$\phb$, that of its {\it bound} states.

\subsection{The scattering component $\phs$}

The scattering component $\phs$ is known from previous 
studies \cite{Bellazzini:2006jb}. One has 
\be
\phs (t,x,i) = \int_{-\infty}^{\infty} \frac{\rd k}{2\pi \sqrt
{2|k|}}
\left[a_i^\ast (k) \e^{\ri (|k|t-kx)} +
a_i (k) \e^{-\ri (|k|t-kx)}\right ] \,  , 
\label{sol1}
\ee
where $\{a_i(k),\, a^*_i(k)\, :\, k\in \RR\}$ generate the 
reflection-transmission (boundary) algebra \cite{Mintchev:2003ue, Liguori:1996xr} 
corresponding to the scattering matrix (\ref{S1}). This is an associative algebra $\mathcal{A}$ 
with identity element $\bf 1$, whose generators $\{a_i(k),\, a^{* i}(k)\, :\, k\in \RR\}$ 
satisfy the commutation relations  
\bea
&a_{i_1}(k_1)\, a_{i_2}(k_2) -  a_{i_2}(k_2)\, a_{i_1}(k_1) = 0\,  ,
\label{ccr1} \\
&a^\ast_{i_1}(k_1)\, a^\ast_{i_2}(k_2) - a^\ast_{i_2}(k_2)\,
a^\ast_{i_1}(k_1) = 0\,  ,
\label{ccr2} \\
&a_{i_1}(k_1)\, a^\ast_{i_2}(k_2) - a^\ast_{i_2}(k_2)\,
a_{i_1}(k_1) = 
2\pi \left [\delta_{i_1 i_2} \delta(k_1-k_2) +
S_{i_1 i_2}(k_1) \delta(k_1+k_2)\right ] {\bf 1}\,  , 
\nonumber \\ 
\label{ccr3}
\eea 
and the constraints\footnote{Eq. (\ref{unit2}) guarantees their consistency.} 
\be
a_i(k) = \sum_{j=1}^n S_{ij} (k) a_j (-k) \, , \qquad 
a^\ast_i (k) = \sum_{j=1}^n a^\ast_ j(-k) S_{ji} (-k)\, .    
\label{constr1}
\ee

Time reversal is implemented in the algebra $\mathcal{A}$ by 
\be 
T a(k) T^{-1} = a(-k)\, , \qquad T a^\ast (k) T^{-1} = a^\ast (-k) \, ,  
\label{timerev2}
\ee 
$T$ being {\it antiunitary}. This action is consistent with 
(\ref{ccr1}-\ref{constr1}) provided that 
\be 
{\overline {S}}(k) = S(-k) \, ,  
\label{real}
\ee
where the bar stands for complex conjugation. Combining (\ref{Ha}) and (\ref{real}), 
one infers\footnote{The superscript $t$ denotes transposition.} 
\be 
S (k)^t = S (k) \, , 
\label{sym}
\ee 
which, in view of (\ref{S2}), implies 
\be 
U^t = U\, .  
\label{condition1}
\ee 
Therefore, the boundary conditions (\ref{bc}) respect time-reversal 
invariance only for symmetric $U$, which will be our choice in what follows. 
Accordingly, the diagonalizing matrix $\U$ in (\ref{d1}) can be chosen to satisfy 
\be 
\U^t = \U^{-1} \, , \qquad {\overline \U} = \U \, . 
\label{d0}
\ee
The case $U^t \not= U$ with broken time reversal has been investigated 
recently in \cite{Bellazzini:2009nk}. 

The field (\ref{sol1}) obviously satisfies the equation of motion (\ref{eqm1}).  
Using (\ref{constr1}), one can show after some algebra that $\phs$ satisfies 
the boundary condition (\ref{bc}) as well. By means of (\ref{ccr1}-\ref{ccr3}) and 
(\ref{sym}) one can also verify that $\phs$ satisfies the canonical relations (\ref{initial1}). 
As far as (\ref{initial2}) is concerned, one obtains 
\be 
[(\prt_t\phs )(0,x_1,i_1)\, ,\, \phs (0,x_2,i_2)] = 
-\ri \delta_{i_1i_2}\, \delta (x_1-x_2) - 
\ri \int_{-\infty}^{\infty} \frac{\rd k}{2\pi } \e^{\ri k\xt_{12}}S_{i_1i_2} (k)\, ,
\label{initial3}
\ee
with $\xt_{12} \equiv x_1+x_2$. The integral in the right hand side of 
(\ref{initial3}) can be computed using the analytic properties of $S(k)$. 
Let us denote by $\cP_+=\{\ri \eta \, :\, \eta>0\}$ and by 
$\cP_-=\{\ri \eta \, :\, \eta<0\}$ the poles of $S(k)$ 
belonging to the upper and lower half-plane respectively. $\cP_+$ 
collects the bound states, whereas $\cP_-$ the antibound states. 
The set $\cP_+$ ($\cP_-$) coincides with the {\it distinct} positive (negative) 
$\eta_i$ appearing in eq. (\ref{p1}). Let us introduce also the matrix 
\be 
R^{(\eta )}_{i_1i_2} = 
\frac{1}{\ri \eta }\, \lim_{k\to \ri \eta } (k-\ri \eta )S_{i_1i_2}(k)\, , \qquad
\ri\eta \in \cP_\pm \, . 
\label{p2}
\ee 
Now, keeping in mind that $\xt_{12} >0$, the integral in (\ref{initial3}) 
can be computed by means of the Cauchy integral formula. One finds 
\be 
[(\prt_t\phs )(0,x_1,i_1)\, ,\, \phs (0,x_2,i_2)] = 
-\ri \delta_{i_1i_2}\, \delta (x_1-x_2) + 
\ri \sum_{\ri\eta \in \cP_+}\eta \e^{-\eta
\xt_{12}} R^{(\eta )}_{i_1i_2} \, , 
\label{initial4}
\ee 
Let us briefly analyze this result. 

If $\cP_+=\emptyset$ (no poles in the upper half-plane) the second term in the right hand side of 
(\ref{initial4}) vanishes and the field $\phs$ satisfies the canonical commutation relations 
(\ref{initial1}, \ref{initial2}). Accordingly, we set in this case $\phb (t,x,i)=0$. 
This is the regime previously investigated in \cite{Bellazzini:2006kh, Bellazzini:2008mn}. 

If $\cP_+ \not= \emptyset$, the scattering matrix $S(k)$ admits bound states. 
The scattering states (represented by the plane
waves in (\ref{sol1}))  do not form a complete set, which leads to the second term in the
right hand side of (\ref{initial4}).  In order to compensate this term, one has to
introduce new boundary degrees of freedom  collected in $\phb (t,x,i)\not= 0$. 
Since the guiding line for constructing $\phb$ is local
commutativity, it is instructive to compute the commutator of $\phs$ at generic
points on the graph. One finds 
\bea 
[\phs (t_1,x_1,i_1)\, ,\, \phs (t_2,x_2,i_2)] = \qquad \qquad \qquad \qquad  \quad \nonumber \\
-\frac{\ri}{4} \left [\varepsilon (t_{12}+x_{12}) + \varepsilon (t_{12}-x_{12})\right ]\delta_{i_1i_2} 
-\frac{\ri}{4} \left [\varepsilon (t_{12}+\xt_{12}) + \varepsilon (t_{12}-\xt_{12})\right ] S_{i_1i_2}(0) \nonumber \\ 
-\frac{\ri}{2} \sum_{\ri \eta \in \cP_+} 
\left [\theta (t_{12}+\xt_{12})\e^{-\eta (t_{12}+\xt_{12})} - 
\theta (-t_{12}+\xt_{12})\e^{\eta (t_{12}-\xt_{12})}\right ] R^{(\eta )}_{i_1i_2} 
\qquad \nonumber \\ 
-\frac{\ri}{2} \sum_{\ri\eta \in \cP_-} 
\left [\theta (t_{12}-\xt_{12})\e^{\eta (t_{12}-\xt_{12})} - 
\theta (-t_{12}-\xt_{12})\e^{-\eta (t_{12}+\xt_{12})}\right ] R^{(\eta )}_{i_1i_2} \, , 
\quad \; \, 
\label{phscomm1}
\eea 
where $\varepsilon$ is the sign function, $t_{12} = t_1-t_2$, 
$x_{12}=x_1-x_2$ and $\xt_{12} = x_1+x_2$. From (\ref{phscomm1}) one gets at 
space-like separations $t_{12}^2-x_{12}^2 <0$ 
\be 
[\phs (t_1,x_1,i_1)\, ,\, \phs (t_2,x_2,i_2)] = 
\ri \sum_{\ri \eta \in \cP_+} \e^{-\eta \xt_{12}} \sinh (\eta t_{12})\, R^{(\eta
)}_{i_1i_2} \, . 
\label{phscomm2}
\ee 
Summarizing, we proved the following statement. 
\bigskip

{\bf Proposition 2}: {\it The scattering component $\phs$ is a canonical local field only in absence of 
boundary bound states ($\cP_+=\emptyset$).} 
\bigskip

In the case $\cP_+\not= \emptyset$ our strategy below will be construct a 
boundary component $\phb$, 
such that the total field $\ph = \phs +\phb$ is {\it both canonical and local}. 

\subsection{The BBS component $\phb$} 

We assume in this subsection $\cP_+ \not= \emptyset$ and introduce the set 
$\I_+ = \{i\, :\, \eta_i > 0\}$, where $\eta_i$ are given by (\ref{p1}). 
The wave functions of the bound states of $K=-\prt_x^2$ 
on $\Gamma$ are then $\{\e^{-\eta_i x}\, :\, i\in \I_+\}$. They are normalizable and 
generate the following set $\{\e^{-\eta_i (x \pm t)}\, :\, i\in \I_+\}$ 
of real solutions of 
the equation of motion (\ref{eqm1}). We associate with 
each index $i\in \I_+$ 
a quantum oscillator. The quantum boundary degrees of freedom 
are described therefore by an algebra $\B$, generated 
by $\{b_i, b^\ast_i\, :\, i\in \I_+\}$, which satisfy 
\bea 
b_{i_1}\, b_{i_2} - b_{i_2}\, b_{i_1} = 0\, , \qquad 
b_{i_1}^\ast \, b_{i_2}^\ast  - b_{i_2}^\ast \, b_{i_1}^\ast = 0\, , 
\label{bccr1}
\\
b_{i_1}\, b_{i_2}^\ast  - b_{i_2}^\ast \, b_{i_1} = 
\delta_{i_1 i_2} \qquad \qquad  
\label{bccr2}
\eea 
and commute with $\{a_i(k),\, a_i^\ast (k)\}$. The field $\phb$, defined by 
\be 
\phb (t,x,i) = \frac{1}{\sqrt{2}} \sum_{j\in \I_+} \U_{ij}\left [\left (b^\ast_j + b_j\right ) \e^{-\eta_j (x+t -t_m)} 
+ \ri \left (b^\ast_j - b_j \right ) \e^{-\eta_j (x-t+t_m)} \right ]\, , 
\label{bfield}
\ee 
satisfies the equation of motion (\ref{eqm1}) and the boundary 
condition (\ref{bc}) by construction. The solution depends on a free parameter $t_m \in \RR$, 
whose physical meaning is clarified in the next section. 

In order to investigate locality, we compute the commutator 
\bea 
[\phb (t_1,x_1,i_1)\, ,\, \phb (t_2,x_2,i_2)] = 
-2 \ri \sum_{j \in \I_+} \e^{-\eta_j \xt_{12}} 
\sinh (\eta_j t_{12})\, \U_{i_1j}\, \U^{-1}_{ji_2} \quad 
\nonumber \\ 
= -2 \ri \sum_{\ri \eta \in \cP_+} \e^{-\eta \xt_{12}} 
\sinh (\eta t_{12}) \sum_{j \in \I_\eta } \U_{i_1j}\, \U^{-1}_{ji_2}
\, , \qquad \qquad \qquad \qquad 
\label{phbcomm0}
\eea 
where $\I_\eta = \{i=1,...,n\, :\, \eta_i=\eta \}$. 
Now, making use of the identity 
\be 
R^{(\eta )}_{i_1i_2} = \frac{1}{\ri \eta }\, \lim_{k\to \ri \eta } 
(k-\ri \eta )\left [\U S_d(k)\U^{-1}\right ]_{i_1i_2} = 
2 \sum_{j\in \I_\eta}\U_{i_1j}\, \U^{-1}_{ji_2} \, , 
\label{p22}
\ee
one finds 
\be 
[\phb (t_1,x_1,i_1)\, ,\, \phb (t_2,x_2,i_2)] = 
-\ri \sum_{\ri\eta \in \cP_+} \e^{-\eta \xt_{12}} 
\sinh (\eta t_{12})\, R^{(\eta )}_{i_1i_2} \, .  
\label{phbcomm}
\ee 
We stress that (\ref{phbcomm}) is $t_m$-independent. 
Combining (\ref{phscomm1}) and (\ref{phbcomm}), one gets for 
the commutator of the total field $\ph = \phs + \phb$ 
\bea 
[\ph (t_1,x_1,i_1)\, ,\, \ph (t_2,x_2,i_2)] = \qquad \qquad \qquad \qquad \qquad \qquad 
\nonumber \\ 
-\frac{\ri}{4} \left [\varepsilon (t_{12}+x_{12}) + \varepsilon
(t_{12}-x_{12})\right ]\delta_{i_1i_2}  -\frac{\ri}{4} \left [\varepsilon (t_{12}+\xt_{12})
+ \varepsilon (t_{12}-\xt_{12})\right ] S_{i_1i_2}(0) 
\qquad \quad \nonumber \\ 
-\frac{\ri}{2} \theta (t_{12}-\xt_{12}) \sum_{\ri \eta \in \cP } \e^{\eta
(t_{12}-\xt_{12})} R^{(\eta )}_{i_1i_2} +\frac{\ri}{2} \theta (-t_{12}-\xt_{12})
\sum_{\ri \eta \in \cP} \e^{-\eta (t_{12}+\xt_{12})} R^{(\eta )}_{i_1i_2} 
\qquad \qquad \;  \; 
\label{phtcomm}
\eea 
where $\cP = \cP_+\cup \cP_-$. Taking into account $x_{1,2}\geq 0$, 
one has at space-like separations $t_{12}^2 - x_{12}^2<0$ that 
$|t_{12}|<|x_{12}|<|{\widetilde x}_{12}|$. 
These last inequalities imply the vanishing of 
(\ref{phtcomm}) at space-like distances, showing that 
$\ph$ is local. The canonical commutation relations 
(\ref{initial1}, \ref{initial2}) hold as well. We thus proved the following statement. 
\bigskip

{\bf Proposition 3.} {\it The boundary component $\phb$ defined above is such that the 
total field $\ph = \phs +\phb$ is both canonical and local for any value of the parameter $t_m\in \RR$.} 
\bigskip 

Let us consider now more closely the time evolution of the system. 
The explicit form of the Hamiltonian $H$ can be deduced from the identity 
\be 
[H\, ,\, \ph (t,x,i)] = -\ri (\partial_t\ph) (t,x,i) \, . 
\label{ham00}
\ee
As expected, 
\be 
H = H^{(s)} + H^{(b)} \, , 
\label{ham0} 
\ee 
where the scattering contribution $H^{(s)}$ has the standard form 
\be 
H^{(s)} = \frac{1}{2}\sum_{i=1}^n\int_{-\infty}^{\infty} \frac{\rd k}{2\pi}
|k|\, a_i^\ast (k) a_i (k)\, . 
\label{ham1}
\ee
Concerning the BBS part $H^{(b)}$, eq.(\ref{ham00}) implies 
\be 
[H^{(b)}\, ,\, b_j+b_j^*] = \ri \eta_j (b_j+b_j^*) \, ,\qquad 
[H^{(b)}\, ,\, b_j-b_j^*] = -\ri \eta (b_j-b_j^*) \, , \quad j\in \I_+\, , 
\label{ham2}
\ee 
which lead to the following expression 
\be   
H^{(b)} = \frac{\ri}{2}\sum_{j\in \I_+} \eta_j 
\left (b_j^2-b_j^{\ast 2}\right )\, .  
\label{ham3}
\ee 
We see that $H^{(b)}$ is not at all the Hamiltonian of the conventional harmonic oscillator, 
but resembles that of the damped oscillator \cite{Celeghini:1991yv}. 
This fact has a relevant impact on the spectral properties of the total Hamiltonian 
and the time-evolution of the whole system, which are discussed in section 3. 

In conclusion, the main lesson from this section is that 
the BBS contribution $\phb$ and the time evolution of the total field $\ph$ 
are fixed up to the parameter $t_m$ by the physical requirement of causality (local commutativity).

\subsection{Dual field, chiral fields and vertex operators}

This subsection is a collection of definitions of the basic structures needed for 
bosonization with BBS on $\Gamma$. The dual field $\phd$, defined by 
\be 
\prt_t \phd (t,x,i) = - \prt_x \ph (t,x,i)\, , \quad  
\prt_x \phd (t,x,i) = - \prt_t \ph (t,x,i)\, .
\label{phd}
\ee 
is given by   
\be 
\phd (t,x,i) = \phds (t,x,i) + \phdb (t,x,i) \, , 
\label{phd1} 
\ee
where 
\be
\phds (t,x,i) = \int_{-\infty}^{\infty} 
\frac{\rd k\, \varepsilon (k)}{2\pi \sqrt {2|k|}} 
\left[a^\ast_i(k) \e^{\ri (|k|t-kx)} +
a_i (k) \e^{-\ri (|k|t-kx)}\right ] \,  ,  
\label{ds}
\ee 
and 
\be 
\phdb (t,x,i) = \frac{1}{\sqrt{2}} 
\sum_{j\in \I_+} \U_{ij}\left [-\left (b^\ast_j + b_j\right ) \e^{-\eta_j
(x+t-t_m)}  + \ri \left (b^\ast_j - b_j \right ) \e^{-\eta_j (x-t+t_m)} \right ]\, , 
\label{db}
\ee 
One easily verifies that $\phd$ is a local field as well. We stress that $\ph$ and $\phd$ are however 
not relatively local, which is fundamental for bosonization. 

The chiral fields are given by 
\be 
\ph_{i, R} (t-x)=\ph(t,x,i)+\phd(t,x,i)\, , \qquad 
\ph_{i,L}(t+x)=\ph(t,x,i)-\phd(t,x,i)\, .   
\label{rlbasis}
\ee 
The BBS contributions to (\ref{rlbasis}) are 
\be 
\ph^{(b)}_{i,R}(t-x) = 
\ri \sqrt{2}\sum_{j\in \I_+} \U_{ij} 
\left (b^\ast_j - b_j \right ) \e^{-\eta_j (x-t+t_m)}
\, , 
\label{rb}
\ee 
\be 
\ph^{(b)}_{i,L}(t+x) = 
\sqrt{2}\sum_{j\in \I_+} \U_{ij} 
\left (b^\ast_j + b_j \right ) \e^{-\eta_j (x+t-t_m)}
\, , 
\label{lb}
\ee
and differently from the scattering parts do not 
oscillate in time, but vanish or diverge exponentially 
in the limit $t\to \pm \infty$. 
This is a first signal that the BBS give origin to a kind of instability 
in the theory, producing a {\it complementary} damping/enhancement of 
$\ph^{(b)}_{i,R}$ and $\ph^{(b)}_{i,L}$ in time. We shall 
characterize this instability more precisely later on at the level of correlation
functions.  
 
Once we have the chiral fields, we can introduce the family of vertex operators 
parametrized by $\zeta = (\sigma , \tau) \in \RR^2$ and defined by 
\be  
\V (t,x,i;\zeta ) = z_i\, \kappa_i\, q(i;\zeta) 
:\exp\left \{\ri \sqrt{\pi}\left [\sigma \ph_{i,R}(t-x) + 
\tau \ph_{i,L}(t+x)\right]\right \}: \, , 
\label{vertex1}
\ee 
where: 

(i) $z_i\in \RR$  are normalization constants which are not essential in what follows;  

(ii) $\kappa_i$ are the so called Klein factors, which satisfy 
\be 
\kappa_{i_1} \kappa_{i_2} + \kappa_{i_2} \kappa_{i_1} = 2 \delta_{i_1i_2} {\bf 1}\ ,
\label{klein}
\ee 
and ensure \cite{Bellazzini:2006kh} that the vertex operators (\ref{vertex1}) 
obey Fermi statistics provided that 
\be 
\sigma^2 -\tau^2 = 2k+1\, , \qquad   k\in \ZZ \, ; 
\label{fermi}
\ee 

(iii) the factor $q(i;\zeta)$ is given by 
\be 
q(i;\zeta) = \exp\left [\ri \sqrt{\pi}
\left (\sigma Q_{i,R} -\tau Q_{i,L}\right )\right ]\, ,  
\label{qfact}
\ee 
where 
\be 
Q_{i,Z} = \frac{1}{4} \int_{-\infty}^{\infty} \rd \xi \, \prt_\xi \ph_{i,Z} (\xi)\, ,  
\qquad Z=R,\, L\, . 
\label{lrcharges}
\ee 
are the chiral charges; 

(iv) finally, $: \cdots :$ denotes the normal product in the algebras $\A$ and $\B$. 

\noindent The BBS contribution to the vertex operators (\ref{vertex1}) is encoded in 
the chiral fields (\ref{rb}, \ref{lb}).

\sect{Correlation functions}

We start by emphasizing that the above analysis of 
the massless scalar field $\ph$ with off-critical boundary conditions on $\Gamma$ 
shows that the theory admits two different regimes, which correspond to   
$\cP_+=\emptyset$ and $\cP_+\not=\emptyset$. Investigating the correlation 
functions, we will show now that these two regimes have very different physical properties. 
The considerations in the previous section are purely algebraic and do not 
depend on the representations of $\A$ and $\B$. However, for the physical applications 
(correlation functions, transition amplitudes, etc.), we must choose a 
representation of each of these algebras. Up to unitary equivalence, the 
choice of the Fock representation $\Fb$ of $\B$ is actually unique. This fact  
is essential for the case $\cP_+\not=\emptyset$. 
In analogy with $\B$, we take below the Fock representation 
$\Fa$ of $\A$ as well. The total Fock space is $\F = \Fa \otimes \Fb$, the vacuum 
being $\Omega = \Omega^{(s)} \otimes \Omega^{(b)}$, where $\Omega^{(s)}$ and 
$\Omega^{(b)}$ are the vacuum states in $\Fa$ and $\Fb$ respectively. For the 
vacuum expectation values in $\F$ we adopt the following short notation 
\be 
\langle \Omega\, , {\cal O}\, \Omega \rangle = 
\langle {\cal O} \rangle_0 \, . 
\label{hl0}
\ee 

Let us discuss now the properties of the Hamiltonian (\ref{ham0}) in this representation, 
focusing on the BBS contribution (\ref{ham3}). Since the ${}^\ast$-operation in $\B$ is 
realized as Hermitian conjugation in $\Fb$, the BBS Hamiltonian $H^{(b)}$ 
is a Hermitian operator. However, as shown in Appendix A, $H^{(b)}$ is {\it not} self-adjoint. 
In fact, its domain\break $\D^{(b)}\subset \Fb$, where (\ref{ham2}) hold, contains eigenstates 
with complex eigenvalues, implying that the time evolution of our system is not unitary. 
A direct proof of this statement is given in the next subsection. 
Analogous behavior, based on complex-valued Hamiltonians, 
has been used recently \cite{Rajeev:2007cp} in the description of dissipative quantum mechanics.

\subsection{The half-line} 

We start our considerations with the half-line $\RR_+$. 
Apart of being relatively simple, this case is fundamental because 
by means of (\ref{d3})  
the derivation of the correlation functions on a 
generic graph $\Gamma$ can be reduced to that on $\RR_+$. 

On $\RR_+$ one has 
\be 
U=\e^{2\ri \alpha}\, , \qquad S(k) = \frac{k+\ri \eta}{k-\ri \eta}\, , 
\qquad \eta = \lambda \tan (\alpha ) 
\label{hl1}
\ee
and the boundary condition (\ref{bc}) takes the familiar form of 
mixed (Robin) condition 
\be 
(\partial_x \ph) (t,0) + \eta \ph (t,0) = 0 \, .  
\label{hl2}
\ee
$S(k)$ has one BBS for $\eta >0$, the associated oscillator being 
generated by $\{b,\, b^\ast \}$. For 
the basic two-point correlators one finds ($\xi_{12}\equiv \xi_1-\xi_2, \; 
{\widetilde \xi}_{12}\equiv \xi_1+\xi_2$): 
\bea 
\langle \ph_{R}(\xi_1) \ph_{R}(\xi_2)\rangle_0 &=& 
u(\mu\, \xi_{12}) + 2 \theta(\eta)  \e^{\eta\, ({\widetilde \xi}_{12}-2t_m)} \, , 
\label{cf1} \\ 
\langle \ph_{L}(\xi_1) \ph_{L}(\xi_2)\rangle_0 &=&
u(\mu\, \xi_{12}) + 2 \theta(\eta) \e^{-\eta\, ({\widetilde \xi}_{12}-2t_m)} \, ,  
\label{cf2} 
\eea
\bea
&\langle \ph_{R}(\xi_1) \ph_{L}(\xi_2)\rangle_0 = 
2\theta(\eta)[v_+(-\eta \xi_{12})  - \ri  \e^{\eta\,\xi_{12}}] +
2\theta(-\eta) v_-(-\eta \xi_{12}) - u(\mu\, \xi_{12}),  
\label{cf3} \\
&\langle \ph_{L}(\xi_1) \ph_{R}(\xi_2)\rangle_0 =  
2\theta(\eta )[v_-(\eta \xi_{12})  + \ri \e^{-\eta\,\xi_{12}}] + 
2\theta(-\eta)v_+(\eta \xi_{12}) - u(\mu\, \xi_{12})\, , \quad 
\label{cf4}
\eea 
with  
\be 
u(\xi) = -\frac{1}{\pi} \ln (\ri\xi + \epsilon ) 
=-\frac{1}{\pi} \ln (|\xi|) -\frac{\ri}{2}\varepsilon (\xi)\, , 
\label{log}
\ee 
\be 
v_\pm (\xi) = -\frac{1}{\pi}\, \e^{-\xi}\, \Ei (\xi \pm i\epsilon ) \, , 
\label{Ei}
\ee
where $\epsilon > 0$ and $\Ei$ is the exponential integral function. Finally, 
$\mu$ is a infrared mass parameter (see e.g. \cite{Liguori:1997vd}). 
The correlators (\ref{cf1}-\ref{cf4}) capture all fundamental properties of the theory 
and clearly show the difference between the two cases $\eta <0$ and $\eta >0$. 

For $\eta <0$ the $S$-matrix (\ref{hl1}) has one antibound state. This case 
has been studied in detail in \cite{Liguori:1997vd}. 
The correlators (\ref{cf1}-\ref{cf4}) are invariant under 
time translations, the energy is conserved and $\ph$ has unitary time evolution. 

The presence of a BBS changes completely the situation for $\eta >0$. 
The exponential terms in (\ref{cf1}-\ref{cf4}), collecting in this case the contribution of the 
BBS on $\RR_+$, have two essential features: 

(i) The factor ${\widetilde \xi}_{12}$ in 
(\ref{cf1}, \ref{cf2}) depends on $t_1+t_2$, which shows that the theory is not invariant 
under time-translations. The origin of this symmetry breaking is the BBS contribution in 
the Hamiltonian (\ref{ham1}), which does not annihilate the vacuum $\Omega$,  
\be 
H \Omega = -\frac{\ri}{2}\, \eta\, b^{\ast 2}\, \Omega \not= 0 \, .  
\label{SSB}
\ee 
Accordingly, for $\eta >0$ the energy of the system is {\it not} conserved, which signals  
a nontrivial energy flow crossing the boundary at $x=0$. We stress that this is a purely boundary effect 
and that the energy momentum tensor in the bulk satisfies the continuity equation. 

(ii) For $t_m\not= 0$ the correlators (\ref{cf1}-\ref{cf4}) are {\it not} invariant under 
the conventional time reversal transformation $t\rightarrow -t$. However, the 
combination of this operation with a specific time translation according to 
\be 
t \rightarrow -t+2t_m\, ,
\label{TT}
\ee
generates a symmetry\footnote{Notice that the mapping (\ref{TT}) 
is actually the reflection on $\RR$ with respect to the point $t_m$.} 
of the correlators (\ref{cf1}-\ref{cf4}). It is implemented by 
\bea 
T\ph_{i,R}(t-x)T^{-1} &=& \ph_{i,L}(-t+2t_m+x)\, , 
\label{TR}\\ 
T\ph_{i,L}(t+x)T^{-1} &=& \ph_{i,R}(-t+2t_m-x)\, ,
\label{TL}
\eea
$T$ being an antiunitary operator which leaves invariant the vacuum, $T\Omega = \Omega$. 

A physical observable, which nicely illustrates the above features 
is the vacuum energy of the system. Let us derive first the vacuum energy density  
of the left and right movers separately. Like in the Casimir effect 
on the graph $\Gamma$ \cite{Bellazzini:2008mn}, this density 
is defined via point-splitting by subtracting the contribution on the line, namely 
\be 
\theta_Z (\xi) = \frac{1}{2} \lim_{\xi_{1,2} \to \xi}  
\left [(\partial \ph_Z)(\xi_1)(\partial \ph_Z)(\xi_2) - 
\langle (\partial \ph_Z)(\xi_1)(\partial \ph_Z)(\xi_2) \rangle_{\rm line} \right ] 
\, , \qquad Z=L,\, R\, .
\label{edo}
\ee 
Now, using (\ref{cf1}, \ref{cf2}) one obtains 
\bea 
\E_L(t+x) &=& \langle \theta_L(t+x) \rangle_0 = 
\eta^2 \e^{-2\eta (t+x-t_m)} \, , 
\label{edl} \\
\E_R(t-x) &=& \langle \theta_R(t-x) \rangle_0 = 
\eta^2 \e^{2\eta (t-x-t_m)} \, . 
\label{edr}
\eea 
For the total vacuum energy one gets 
\begin{equation} 
E(t) \equiv \int_0^\infty \rd x [\E_L(t+x)+\E_R(t-x)] = 
\eta \cosh [2\eta (t-t_m)]\, .   
\label{tve}
\end{equation} 
As expected, the vacuum energy is time dependent: it decreases (dissipation) in the interval $(-\infty, t_m)$ 
and increases (enhancement) in $(t_m, \infty)$. $E(t)$ has an absolute minimum at $t=t_m$, which fixes the physical 
interpretation of the free parameter $t_m$ present in the general solution (\ref{bfield}, \ref{db}). 
Obviously, $t_m$ indicates also the instant in which the vacuum energy flow trough the boundary 
changes its direction. Finally, being a reflection with respect to $t_m$, the 
time reversal transformation (\ref{TT}) inverts the regimes of dissipation ($t<t_m$) 
and enhancement ($t>t_m$) of the vacuum energy. 

In order to understand better the above phenomena, it is instructive to 
study the time evolution of the vacuum state $\Omega$. Using the results of 
Appendix A, one can compute the expectation value 
\be 
\langle \Omega\, ,\, \e^{\ri t H}\Omega \rangle = 
\langle \e^{\ri t H} \rangle_0 = \frac{1}{\sqrt{\cosh (\eta t)}} \, , 
\label{vte}
\ee   
which vanishes for $t\to \pm \infty$. In these limits therefore 
$\Omega$ evolves in a state which is orthogonal to itself, which is 
essentially the result of \cite{Celeghini:1991yv} for the 
{\it damped oscillator} with $\eta$ playing the role of {\it friction}. 
In this respect the vacuum energy densities (\ref{edr}, \ref{edl}) 
show the presence of {\it positive} friction in the left sector 
and {\it negative} friction in the right one. 

The existence of friction suggest also the violation of unitarity. In fact, 
deriving along the above lines the time evolution of the one-particle BBS state 
$b^* \Omega$, one finds 
\be 
\langle \e^{\ri t H} b^* \Omega\, ,\, \e^{\ri t H} b^* \Omega \rangle = \cosh (\eta t) \, . 
\label{unitaryloss}
\ee 
The time dependence of (\ref{unitaryloss}) implies the breakdown of unitarity. 

It is interesting at this point to investigate the time evolution of a 
generic coherent state 
\be 
\Omega_c (z) = \e^{zb^\ast -{\overline z}b} \Omega \, , 
\qquad z = x +\ri y \in \CC\, , 
\label{cstate1}
\ee
generated by the boundary degrees of freedom. The main points 
of the computation are sketched in the Appendix A. The result is 
\be 
\langle \Omega_c(z)\, ,\, \e^{\ri t H}\Omega_c(z) \rangle = 
\frac{\e^{-(x^2+y^2) [1+{\rm sech} (\eta t)] - 
2\ri xy \tanh (\eta t)}}{\sqrt{\cosh (\eta t)}}\, , 
\label{cstate2}
\ee
showing that the vacuum $\Omega$ and the coherent state $\Omega_c(z)$ 
behave in the same way in the limit $t \to \pm \infty$. 

The results of this subsection are collected in the following proposition. 
\bigskip 

{\bf Proposition 4:} {\it The boundary bound state, which occurs for 
$\eta>0$ in the massless local scalar field $\ph$ on the half-line $\RR_+$, implies: 

(a) breaking of time translation invariance; 

(b) vacuum decay like in the case of the damped oscillator with friction $\eta$;

(c) nonunitary time evolution;} 
\bigskip 

In agreement with point (a), the dynamics of the system depends on the parameter $t_m$ 
and exhibits two different regimes: the vacuum energy flow trough the boundary is 
outgoing for $t<t_m$ and incoming for $t>t_m$. For simplicity we analyzed in this section the half-line, 
but we will show below that the above results hold for a generic star graph as well.

\subsection{Generic star graph} 

The two-point correlation functions for a general 
star graph $\Gamma$ are easily obtained combining the results on 
$\RR_+$ with the diagonalized form (\ref{d3}) of the $S$-matrix. 
One finds  
\bea 
\langle  \ph_{i_1,R}(\xi_1) \ph_{i_2,R}(\xi_2)\rangle_0 &=&  
\sum_{\ri \eta\in \cP_+}\e^{\eta\, ({\widetilde \xi}_{12}-2t_m)} R_{i_1i_2}^{(\eta)} 
+\delta_{i_1i_2}\, u(\mu\, \xi_{12})\, , 
\label{cf5} \\ 
\langle \ph_{i_1,L}(\xi_1) \ph_{i_2,L}(\xi_2)\rangle_0 &=&
\sum_{\ri \eta\in \cP_+}\e^{-\eta\, ({\widetilde \xi}_{12}-2t_m)} R_{i_1i_2}^{(\eta)}
+ \delta_{i_1i_2}\, u(\mu\, \xi_{12})\, ,  
\label{cf6} 
\eea
\bea
\langle \ph_{i_1,R}(\xi_1) \ph_{i_2,L}(\xi_2)\rangle_0 = 
\qquad \qquad \qquad \qquad \qquad \nonumber \\  
\sum_{\ri \eta\in \cP_+}\left [v_+(-\eta \xi_{12})  - 
\ri \e^{\eta\,\xi_{12}}\right ]R_{i_1i_2}^{(\eta)} +  
\sum_{\ri \eta\in \cP_-} v_-(-\eta \xi_{12})R_{i_1i_2}^{(\eta)} 
- \delta_{i_1i_2}\, u(\mu\, \xi_{12})\, ,  
\label{cf7} \\
\langle \ph_{i_1,L}(\xi_1) \ph_{i_2,R}(\xi_2)\rangle_0 = 
\qquad \qquad \qquad \qquad \qquad \nonumber \\
\sum_{\ri \eta\in \cP_+}\left [v_-(\eta \xi_{12})  + 
\ri \e^{-\eta\,\xi_{12}}\right ]R_{i_1i_2}^{(\eta)} +  
\sum_{\ri \eta\in \cP_-} v_+(\eta \xi_{12})R_{i_1i_2}^{(\eta)} 
- \delta_{i_1i_2}\, u(\mu\, \xi_{12})\, . \quad  
\label{cf8}
\eea 
In (\ref{cf5}, \ref{cf6}) we recognize a $t_1+t_2$-dependence.  
As expected, the presence of BBS ($\cP_+\not= \emptyset$) implies 
the breakdown of  time-translation invariance. Instead of (\ref{vte}), one has on $\Gamma$  
\be 
\langle \Omega\, ,\, \e^{\ri t H}\Omega \rangle = 
\langle \e^{\ri t H} \rangle_0 = 
\left [\frac{1}{\sqrt{\cosh (\eta t)}}\right ]^{n_+}
\, , 
\label{vteg}
\ee 
$n_+$ being the number of positive $\eta_i$. Equation 
(\ref{vteg}) shows the time evolution of the vacuum. For the 
vacuum energy density of the chiral fields on $\Gamma$, one gets from 
(\ref{cf5}, \ref{cf6})  
\be 
\E_{i,L}(t+x) = 
\sum_{\ri \eta \in \cP_+} \eta^2 \e^{-2\eta (t+x-t_m)} R_{ii}^{(\eta)} \, , 
\qquad \E_{i,R}(t-x) = 
\sum_{\ri \eta \in \cP_+} \eta^2 \e^{2\eta (t-x-t_m)} R_{ii}^{(\eta)}\, , 
\label{edlr}
\ee 
which behave in much the same way as on the half-line. Finally, the total vacuum energy is 
\begin{equation}
E(t) = \sum_{\ri \eta \in \cP_+} \eta R_{ii}^{(\eta)} \cosh [2\eta(t-t_m)] \, . 
\label{tvegamma}
\end{equation} 

In conclusion, the fields $\ph_{i,Z}$ on $\Gamma$ have the same features as their 
counterparts on the half-line. Proposition 4 is therefore valid on any star graph.

\sect{Luttinger liquid with off-critical boundary conditions} 

The quantum field theory on $\Gamma$, receiving much attention \cite{NFLL}-\cite{BCM}, is the  
Tomonaga-Luttinger (TL) model because it captures the universal features of a 
wide class of one-dimensional quantum many-body systems, called Luttinger liquids. 

The dynamics of the TL model is defined by the Lagrangian density  
\begin{equation} 
\mathcal{L} = \ri \psi_1^*(\prt_t + v_F\prt_x)\psi_1 +  \ri \psi_2^*(\prt_t - v_F\prt_x)\psi_2 
-g_+(\psi_1^* \psi_1+\psi_2^* \psi_2)^2 - g_-(\psi_1^* \psi_1-\psi_2^* \psi_2)^2\, ,  
\label{lagrangian}
\end{equation} 
where $\psi_\alpha(t,x,i)$ with $\alpha =1,2$ are complex fermion fields, 
$v_F$ is the Fermi velocity and $g_\pm \in \RR$ are the coupling constants\footnote{Another 
frequently used notation is $g_2=g_+-g_-$ and $g_4=g_+ + g_-$.}. 
$\mathcal L$ is invariant under the phase transformations  
\begin{eqnarray}
\psi_\alpha &\rightarrow& \e^{\ri s} \psi_\alpha \, , \quad \qquad \; \, 
\psi^*_\alpha \rightarrow \e^{-\ri s} \psi^*\, ,  \qquad \quad \; \, s \in \RR\, ,
\label{V}\\
\psi_\alpha &\rightarrow& \e^{-\ri(-1)^\alpha {\tilde s}} \psi_\alpha  \, , \quad 
\psi^*_\alpha \rightarrow \e^{\ri(-1)^\alpha {\tilde s}} \psi^*_\alpha\, , 
\qquad {\tilde s} \in \RR\, , 
\label{A}
\end{eqnarray} 
showing that $U(1)\otimes {\widetilde U}(1)$ is the (internal) symmetry group of the model. 

It is well known (see e. g. \cite{Hald,Voit}) that the TL model on the line 
$\RR$ is exactly solvable by bosonization, the solution being expressed in 
terms of the massless fields $\ph$ and $\phd$. 
On the graph $\Gamma$ the situation is a bit more involved because of 
the additional interaction localized in the vertex $V$. 
At criticality, the solution via bosonization has been given in 
\cite{Bellazzini:2006kh, Bellazzini:2008mn}. 
The basic physical observable, derived there, 
is the electromagnetic conductance matrix 
\be 
G_{ij} = G_{\rm line}\left (\delta_{ij}  - S_{ij} \right ) \, , 
\label{cond1}
\ee 
where $G_{\rm line}$ is the conductance on the line and $S$ is the scattering 
matrix which is constant because of scale invariance. 
Our main goal below is to derive the counterpart of (\ref{cond1}) away from 
criticality. For this purpose it is enough 
to consider the TL model in the special case 
when $g_+=-g_-\equiv g\pi >0$ and $v_F=1$. The classical equations of motion 
of this system, known as the Thirring model \cite{Thirr}, can be written in 
a simple matrix form 
\be 
\ri (\gamma_t \prt_t - \gamma_x \prt_x)\psi (t,x,i) = 
2\pi g [\gamma_t J_t(t,x,i) - \gamma_x J_x(t,x,i)]\psi (t,x,i) \, , 
\label{eqmt} 
\ee 
where
\be 
\psi (t,x,i)=\pmatrix{ \psi_1(t,x,i) \cr \psi_2(t,x,i) \cr}\, ,  
\qquad 
\gamma_t = \pmatrix{ 0 & 1 \cr 1 & 0 \cr}\, , \qquad
\gamma_x = \pmatrix{ 0 & 1 \cr -1 & 0 \cr} \, .
\label{gamma}
\ee 
and 
\be 
J_\nu (t,x,i) = \overline \psi (t,x,i) \gamma_\nu \psi (t,x,i) \, , \qquad 
\overline \psi \equiv \psi^\ast \gamma_t \, . 
\label{currt1} 
\ee
is the conserved $U(1)$-current. The ${\widetilde U}(1)$-counterpart is 
\be
\Jt_\nu (t,x,i) = \overline \psi (t,x,i) \gamma_\nu \gamma_5 \psi (t,x,i) \, , \qquad 
\gamma_5=-\gamma_t \gamma_x \, , 
\label{dcurrt1}  
\ee 
which is also conserved.

\subsection{Bosonization} 

{}For solving the Thirring model, we set $\sigma >0$ and 
\be
\psi_1 (t,x,i) = \frac{1}{\sqrt {2\pi}}\mathcal{V} (t,x,i;\zeta )\, , \qquad 
\psi_2 (t,x,i) = \frac{1}{\sqrt {2\pi}}\mathcal{V} (t,x;\zeta^\prime ) \, ,
\label{psit}
\ee 
the vertex operators being defined by (\ref{vertex1}) with $\zeta = (\sigma, \tau)$ 
and $\zeta^\prime = (\tau, \sigma)$. In order to have canonical fermions we require 
\be 
\sigma^2-\tau^2=1\, ,  
\label{condt}
\ee
implying $\sigma \not= \pm \tau$. We consider here the standard fermionic Luttinger 
liquid on $\Gamma$, but the results below have a straightforward extension to the 
anyonic case \cite{BCM} as well.  

The quantum current $J_\nu $ is constructed by point-splitting according to 
\be
J_\nu (t,x,i) = \frac{1}{2} \lim_{\epsilon \to +0} Z (\epsilon) 
\left [\, \overline \psi (t,x,i)\gamma_\nu \psi (t, x+\epsilon, i) + 
\overline \psi (t, x+\epsilon, i) \gamma_\nu \psi (t, x, i)\, \right ]\, , 
\label{pointsplit3}  
\ee 
where $Z (\epsilon)$ is some renormalization constant. The latter can 
be fixed \cite{Bellazzini:2006kh} in such a way that 
\be 
J_\nu (t,x,i) = - \frac{1}{(\sigma + \tau)\sqrt \pi}\,  \prt_\nu \ph (t,x,i) \, , 
\label{currt2} 
\ee
thus generating the $U(1)$-phase transformations (\ref{V})  
\be 
[J_t (t,x,i)\, ,\, \psi(t,y,j)] = -\delta(x-y) \delta_{ij} \psi (t,x,j) \, . 
\label{wi1}
\ee 
Because of (\ref{currt2}) the quantum equation of motion takes the form 
\be
i(\gamma_t \prt_t - \gamma_x \prt_x)\psi (t,x,i) = 
-\frac{2g\sqrt \pi }{(\sigma + \tau)} : \left (\gamma_t \prt_t \ph - 
\gamma_x \prt_x \ph \right ) \psi : (t,x,i)  \, .  
\label{qeqmt}
\ee
Now, using the vertex realization (\ref{psit}) of $\psi$, one easily 
verifies that (\ref{qeqmt}) is satisfied provided that 
\be 
\tau(\sigma+\tau) = g\, . 
\label{eqmts}
\ee
Combining (\ref{condt}) and (\ref{eqmts}), one determines $\sigma$ and $\tau$ in terms of the coupling 
constant  
\be 
\sigma = \frac{1+g}{\sqrt {1+2g}} >0\, , \qquad \tau =  \frac{g}{\sqrt {1+2g}} \, , 
\label{solt}
\ee 
which fix the solution completely. 

The quantum current $\Jt_\nu$ is defined by point splitting as well. One has 
\be
\Jt_\nu (t,x,i) = \frac{1}{2} \lim_{\epsilon \to +0} {\widetilde Z} (\epsilon) 
\left [\, \overline \psi (t,x,i)\gamma_\nu \gamma_5 \psi (t, x+\epsilon, i) + 
\overline \psi (t, x+\epsilon, i) \gamma_\nu \gamma_5 \psi (t, x, i)\, \right ] 
\label{pointsplit4}  
\ee 
with $\gamma_5=-\gamma_t \gamma_x$. A suitable choice \cite{Bellazzini:2006kh} of the renormalization constant 
${\widetilde Z} (\epsilon)$ leads to 
\be 
\Jt_\nu (t,x,i) = -\frac{1}{(\sigma - \tau)\sqrt \pi}\,  \prt_\nu \phd (t,x,i) \, , 
\label{currt3} 
\ee
which is indeed conserved and generates the 
${\widetilde U}(1)$-phase transformations (\ref{A}) according to, 
\be 
[\Jt_t (t,x,i)\, ,\, \psi_\alpha (t,y,j)] = 
\left\{\begin{array}{cc}
-\delta(x-y) \delta_{ij} \psi_1 (t,y,j) \, ,
& \quad \mbox{$\alpha =1$}\, ,\\[1ex]
\; \; \delta(x-y) \delta_{ij} \psi_2 (t,y,j)\, ,
& \quad \; \mbox{$\alpha = 2$}\, . \\[1ex]
\end{array} \right. 
\label{wi2}
\ee 
It is worth mentioning that the boundary condition (\ref{bc}) is transferred to a {\it local} condition on 
$\Jt_\nu$, even away from criticality. Indeed, (\ref{bc}) implies 
\be 
(\partial_x \Jt_x) (t,0,i) + \lambda \sum_{j=1}^n B_{ij} \Jt_x (t,0,j) = 0 \, , 
\label{lbc}
\ee
where 
\be 
B=\U B_d\, \U^{-1}\, , \qquad B_d = {\rm diag}\left (\tan(\alpha_1), \tan(\alpha_2),...,\tan(\alpha_n)\right ) \, . 
\nonumber 
\ee
A simple way of proving eq. (\ref{lbc}) is to check it directly on the half-line ($n=1$) and extend the result on 
$\Gamma$ by means of (\ref{d1}-\ref{p1}). Using the duality relation (\ref{phd}), the boundary condition 
(\ref{lbc}) can be written also in terms of the $U(1)$-current 
\be 
(\partial_x J_t) (t,0,i) + \lambda \sum_{j=1}^n B_{ij} J_t (t,0,j) = 0 \, . 
\label{lbc1}
\ee

We turn now to the important issue of symmetries.  
Clearly, the boundary conditions (\ref{bc}) influence the symmetry content of the model on $\Gamma$. 
The relation between symmetries and boundary conditions is implemented by the 
Kirchhoff's rule, which must be imposed in the vertex of $\Gamma$ on any conserved current 
in order to generate a time-independent charge from it. The matrix $U$ in (\ref{bc}) for instance 
parametrizes all boundary conditions which ensure the Kirchhoff rule for the energy-momentum tensor 
of $\ph$ and thus the time-independence of the relative Hamiltonian. Since different conserved currents generate 
in general nonequivalent Kirchhoff's rules, one may expect the presence of obstructions for 
lifting all symmetries on the line to symmetries on $\Gamma$. 
It may happen in fact that 
two Kirchhoff's rules are in contradiction for {\it generic}\footnote{Excluding some exceptional  
boundary conditions in graphs with even number $n=2m$ of edges for which the system 
behaves as a bunch of $m$ independent lines.} boundary conditions. In this case one 
can preserve on $\Gamma$ one of the corresponding symmetries, but not both of 
them. This is actually the case with the $U(1)\otimes{\widetilde U}(1)$-group. 

Let us consider first the $U(1)$-factor. Using (\ref{dec1}), the associated current (\ref{currt2}) 
can be decomposed in scattering and BBS part, namely 
\be 
J_\nu (t,x,i) =  J_\nu^{(s)}(t,x,i) + J_\nu^{(b)}(t,x,i)  \, . 
\label{currd1} 
\ee
According to \cite{Bellazzini:2006kh} 
\be 
\sum_{i=1}^{n}J^{(s)}_x(t,0,i)=0\quad \Longleftrightarrow \; \sum_{i=1}^{n}S_{ij}(k)=1 \; 
\Longleftrightarrow \; \sum_{i=1}^{n}U_{ij}=1\, , \quad  \forall j =1,...,n\, .  
\label{kirchV} 
\ee
For the BBS contribution one gets from (\ref{bfield}) 
\be 
\sum_{i=1}^{n}J^{(b)}_x(t,0,i)=0 \quad 
\Longleftrightarrow \; \sum_{i=1}^{n}\U_{ij}=0 \, , \quad \forall j \in \I_+\, .  
\label{kirchVb} 
\ee
In Appendix B we show that the constraint (\ref{kirchV}) on $U$ actually implies the 
condition (\ref{kirchVb}) on $\U$. Therefore the $U(1)$-Kirchhoff rule holds if and only if 
the matrix $U$ satisfies (\ref{kirchV}). Analogously, for the 
${\widetilde U}(1)$-Kirchhoff rule one gets  
\be 
\sum_{i=1}^{n}\Jt_x(t,0,i)=0  \;  \Longleftrightarrow \sum_{i=1}^{n}U_{ij}=-1\, ,
\label{kirchA}
\ee 

Comparing the constraints (\ref{kirchV}) and (\ref{kirchA}) on the matrix $U$, we see that 
they cannot be satisfied simultaneously. Therefore, differently from the line, 
on $\Gamma$ one must choose 
between the $U(1)$-symmetry or its dual ${\widetilde U}(1)$. The corresponding systems have 
{\it different} physical 
properties. Coupling minimally the model to an external electromagnetic field $A_\nu (t,x,i)$, 
one finds that (\ref{kirchV}) implies electric charge conservation. On the other hand, imposing 
(\ref{kirchA}), the electric current $J_x$ does not satisfy the Kirchhoff rule and 
the electric charge is no longer conserved. However, equation (\ref{kirchA}) implies by duality (\ref{phd}) 
that the charge {\it density} satisfies 
\be 
\sum_{i=1}^{n}J_t(t,0,i)=0 \, , 
\label{kirchD}
\ee
which is considered \cite{DR} to be the characteristic feature of a {\it super-conducting} junction. 
Both (\ref{kirchV}) and (\ref{kirchA}) have therefore a precise physical interpretation.

\subsection{Conductance}

In order to derive the electric conductance tensor $G_{ij}$, we couple the system to a {\it classical} 
external field $A_\nu (t,x,i)$ by means of the substitution 
\be 
\prt_\nu \longmapsto \prt_\nu + \ri A_\nu (t,x,i)  
\label{covder}
\ee 
in eq. (\ref{eqmt}). The resulting Hamiltonian is {\it time dependent} and the conductance can be extracted 
from the linear term of the expansion of $\langle J_x(t,x,i)\rangle_{A_\nu}$ in terms of $A_\nu$. 
This term can be computed by linear response theory \cite{FW}. Referring for the details 
to \cite{Bellazzini:2006kh}, one has  
\bea
\langle J_x(t,x,i)\rangle_{A_\nu} =  
\langle J_x(t,x,i) \rangle +\ri\int_{-\infty}^{t} \rd\tau
\langle [H_{\rm int}(\tau)\, ,\, J_x(t,x,i)]\rangle = \qquad \qquad 
\nonumber \\
\frac{1}{\pi(1+2g)}\left [A_x(t,x,i)+
\ri \sum_{j=1}^{n}\int_{-\infty}^{t} \rd\tau 
\int_0^\infty \rd y A_y(\tau,y,j)
\langle [\partial_y\ph(\tau,y,j)\, ,\, \partial_x\ph(t,x,i)]\rangle \right ]\, . \nonumber 
\eea 
\be
{}\label{lrt1}
\ee
Let us consider now a uniform electric field\footnote{We use the Weyl gauge $A_t=0$.} 
$E(t,i)=\partial_t A_x(t,i)$ which is switched on at $t=t_0$, i.e. $A_x(t,i) = 0$ for $t<t_0$. 
Using the Fourier transform of (\ref{phtcomm}), one can compute the 
expectation value (\ref{lrt1}). The final result is 
\bea
\langle J_x(t,0,i)\rangle_{A_\nu} = \qquad \qquad \qquad \qquad \qquad \qquad 
\nonumber \\
G_{\rm line}\sum_{j=1}^n\int_{-\infty}^\infty \frac{\rd \omega}{2\pi} 
{\hat A}_x(\omega, j)\e^{-\ri \omega t} 
\left [\delta_{ji} - S_{ji}(\omega ) - 
\sum_{\ri \eta \in \cP} R_{ji}^{(\eta)} \frac{\eta}{\eta + \ri \omega} 
\e^{(t-t_0)(\eta+\ri \omega)} \right ]\, , 
\label{lrt2}
\eea 
where 
\be 
G_{\rm line} = \frac{1}{2\pi(1+2g)}\, , \qquad 
{\hat A}_x(\omega, i) = \int_{-\infty}^\infty \rd \tau\, \e^{\ri \omega \tau} A_x(\tau,i)\, .  
\ee 
The conductance can be extracted directly from (\ref{lrt2}) and reads 
\be 
G_{ij}(\omega, t-t_0) =  
G_{\rm line} \left [\delta_{ij} - S_{ij}(\omega ) - \e^{\ri (t-t_0)\omega}
\sum_{\ri \eta \in \cP} \frac{\eta \e^{(t-t_0)\eta}}{\eta + \ri \omega} R_{ij}^{(\eta)} \right ]\, , \quad t > t_0\, , 
\label{cond2}
\ee 
which is the off-critical generalization of (\ref{cond1}). Notice that $G_{ij}(\omega, t-t_0)$ is in 
general complex, which leads to a nontrivial impedance \cite{Bellazzini:2008mn}. 
The result (\ref{cond2}) concerns the Thirring model, 
but the TL-conductance tensor is essentially the same, the only 
change being in the form of $G_{\rm line}$ which now depends \cite{BCM} 
on $g_\pm$ and $v_F$. 

It is worth mentioning that (\ref{cond2}) satisfies the Kirchhoff rule 
\be 
\sum_{j=1}^n G_{ij}(\omega, t-t_0) = 0\, , 
\label{kirchG}
\ee 
provided that the electric charge is conserved, i.e. (\ref{kirchV}) holds. 
In fact, (\ref{kirchV}) ensures the compensation of the 
first two terms in (\ref{cond2}). As far as the $R_{ij}^{(\eta)}$-contribution is concerned, 
combining (\ref{kirchV}) with the definition (\ref{p2}) one gets  
\be 
\sum_{j=1}^n R^{(\eta)}_{ij} = 0\, , \qquad \forall \, \eta \in \cP\, ,  
\label{kirchR}
\ee 
which completes the argument. 

The off-critical conductance (\ref{cond2}) keeps track of the two regimes 
$\cP_+=\emptyset$ and $\cP_+\not=\emptyset$ of the theory. In fact, it is 
instructive to consider them separately. 

(i) $\cP_+=\emptyset$: In this case the sum in the right hand side of (\ref{cond2}) runs 
only over the antibound states $\cP_-$, all of them producing damped oscillations in $t-t_0$. 
Therefore  
\be 
\lim_{t\to \infty} G_{ij}(\omega, t-t_0) = 
\lim_{t_o\to -\infty} G_{ij}(\omega, t-t_0) = 
G_{\rm line} \left [\delta_{ij} - S_{ij}\left (\omega \right )\right ] \, ,  
\label{limits}
\ee 
which gives the conductance, one will observe in this regime, long time after 
switching on the external field. At a critical point $S$ is constant and (\ref{limits}) 
reproduces (\ref{cond1}), which is claimed \cite{NFLL, Oshikawa:2005fh, Aff, Chamon1} to 
describe some universal features of Luttinger junctions. 

(ii) $\cP_+\not=\emptyset$: Now the sum in the right hand side of (\ref{cond2}) involves 
terms with $\eta>0$, which give origin to oscillations whose amplitude is growing 
exponentially with $t-t_0$.

It is worth stressing that presently we are not aware if the response of real-life quantum wire junctions to an external 
electric field can be of the type (i) and/or (ii). It is nice however that the predicted behavior of the 
conductance in these two cases is quantitatively different, thus providing a clear experimental signature. 

The result (\ref{cond2}) can be applied also at the level of effective theory. 
An example, which frequently appears in the literature \cite{Oshikawa:2005fh, Aff, Chamon1}, 
is the Y-junction with the geometry displayed in Fig. 2. One has three external half lines $E_i$ 
and a ring composed of three compact internal edges $I_i$. A magnetic flux $\phi$ is 
crossing the ring. The complete field theory analysis of the Luttinger liquid 
on a graph with this geometry is very complicated problem. 
One approximate way to face this problem could 
be to use the star product approach \cite{KS} or 
the ``gluing" technique \cite{Mintchev:2007qt, Ragoucy:2009hf} for 
deriving the $3\times 3$ scattering matrix relative to the {\it external} edges. 
This $S$-matrix can be used for developing a simplified model 
with {\it one effective} off-critical junction for which (\ref{cond2}) applies.  

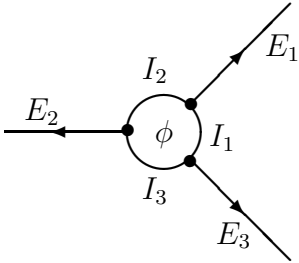
\begin{figure}[tb]
\setlength{\unitlength}{1mm}
\begin{picture}(20,20)(-42,20)
\put(25.2,1.6){\makebox(20,20)[t]{$\phi$}}
\put(42,11){\makebox(18,22)[t]{$E_1$}}
\put(9,3.5){\makebox(20,21)[t]{$E_2$}}
\put(34.5,-12){\makebox(20,20)[t]{$E_3$}}
\put(33,1){\makebox(20,20)[t]{$I_1$}}
\put(24,10){\makebox(20,20)[t]{$I_2$}}
\put(24,-6){\makebox(20,20)[t]{$I_3$}}
\put(20.4,1){\makebox(20,20)[t]{$\bullet$}}
\put(28.8,4.5){\makebox(20,20)[t]{$\bullet$}}
\put(28.7,-3){\makebox(20,20)[t]{$\bullet$}}
\thicklines 
\put(35.2,20){\circle{10}}
\put(39,24){\line(1,1){13}}
\put(30,20){\line(-1,0){16}}
\put(39,16){\line(1,-1){13}}
%
%
\put(46,31){\vector(1,1){0}}
\put(46,9){\vector(1,-1){0}}
\put(20,20){\vector(-1,0){0}}
%
\end{picture} 
\vskip 2truecm
\caption{ A graph with 3 external and 3 internal edges.} 
\label{graph}
\end{figure}

In conclusion, we emphasize that the results of this subsection are obtained 
in an abstract setting, where we attempt to describe the physics of a quantum wire junction  
by a Luttinger liquid with specific off-critical boundary conditions at the vertex of a star graph, 
approximating the junction. Further investigations are needed for clarifying both 
the plausibility of these assumptions and the applicability of the result (\ref{cond2}) to 
realistic quantum wire junctions.

\sect{Outlook and conclusions} 

We started this paper by developing the theory of a massless scalar field $\ph$ with 
off-critical boundary conditions on a star graph. The vertex of the graph is represented 
as a point-like defect characterized by a scattering matrix $S(k)$. 
In general $S(k)$ admits bound and antibound states. This fact determines two 
physically different regimes of the theory. In absence of bound states 
the field $\ph$ has unitary time evolution and the energy is conserved. 
The behavior of the system changes completely if bound states are present.  
Each such a state generates an oscillator in the spectrum of $\ph$, whose contribution 
is fixed by local commutativity up to a common free parameter $t_m$. 
These oscillators break the invariance under time translations 
and drive the system out of equilibrium. Accordingly, 
the vacuum energy is time dependent: it decays exponentially for $t<t_m$ 
and grows at the same rate for $t>t_m$. The two regimes are related by time reversal 
and correspond to a nontrivial outgoing and incoming vacuum energy flows. 
Recalling that the traditional way \cite{CL} for driving a system out of equilibrium is to couple it 
with a ``bath" of external oscillators, we discovered above that such oscillators 
are automatically present for $S(k)$ with bound states. Such boundary conditions provide 
therefore an intrinsic mechanism for constructing non-equilibrium quantum systems. 
It is worth stressing that this mechanism is based on purely boundary effects and is 
completely fixed by the fundamental physical requirement of causality. 

It turns out that the behavior of the Luttinger liquid on $\Gamma$ 
is affected by the above phenomena. This is true in particular for 
the electromagnetic conductance $G$. Deriving an explicit formula 
for $G$, we have shown that it depends on the time elapsed after switching 
on the external electric field and develops oscillations with both exponentially growing 
and decaying amplitudes, corresponding to the bound and antibound states respectively. 

Being interested in the Luttinger liquid, we investigated the massless scalar field on 
$\Gamma$, but the BBS phenomena described in this paper take place also in the 
massive case $m>0$, provided that the boundary parameter 
$\eta$ satisfies $\eta > m$. In the range $0<\eta<m$ the BBS do not 
produce instabilities \cite{Mintchev:2004jy}. 
  
Let us comment finally on some possible generalizations of the above results. 
It will be useful for the applications to extend our framework to boundary conditions 
which violate time reversal \cite{BSMS, Chamon1}. This more technical then conceptual 
problem has been addressed recently within the above framework in \cite{Bellazzini:2009nk}. 
The study of the rich spectrum \cite{Ines} of effects away from equilibrium is essential for the experiment. 
More general systems of wires with several junctions attract also much attention \cite{KS}-\cite{Sch}. 
Networks involving compact edges, like that in Fig. 2, are expected to have richer BBS content and 
represent a challenging open problem. 

We stress that the phenomena, investigated in this paper, illustrate actually very general features 
of systems with boundaries. In fact, the BBS mechanism, described above, has a straightforward 
extension \cite{BMS} from point like to higher dimensional boundaries 
and admits therefore other potential applications.

\bigskip

\noindent{\bf Acknowledgments} 
\bigskip 

The research of B. B. has been supported in part by the NSF grant number NSF-PHY-0757868. 

\appendix

\sect{BBS Hamiltonian and vacuum decay} 

We demonstrate first that the BBS Hamiltonian (\ref{ham2}) is Hermitian but not self-adjoint. 
It is enough to consider the case of one BBS, i.e. 
\be 
H^{(b)} =  \frac{\ri}{2} \eta \left (b^2 - b^{\ast 2}\right )\, , \qquad \eta >0\, .  
\label{A1}
\ee 
This Hamiltonian generates the correct time evolution 
of $\phb$, provided that $H^{(b)}$ and $b\pm b^\ast$ 
have a common and invariant domain $\D^{(b)}$ where the 
commutation relations 
\be 
[H^{(b)}\, ,\, b+b^*] = \ri \eta (b+b^*) \, ,\qquad 
[H^{(b)}\, ,\, b-b^*] = -\ri \eta (b-b^*) \, , 
\label{A2}
\ee 
hold. We show now that $H^{(b)}$ is Hermitian but not self-adjoint on $\D^{(b)}$. 
In fact, let $\Phi \in \D^{(b)}$ be an eigenstate 
of $H^{(b)}$ with eigenvalue $E$, i.e. $H^{(b)}\Phi = E\Phi$. Suppose that $E\in \RR$, otherwise 
there is nothing to prove. Using (\ref{A2}) one easily verifies that 
$(b+b^*) \Phi \in \D^{(b)}$ is an eigenvector of $H^{(b)}$ with eigenvalue $E+\ri \eta$, 
which is complex since $\eta>0$. This concludes the argument. 

Our next step is to sketch the derivation of equations (\ref{vte}) and (\ref{cstate2}). We start with 
(\ref{vte}), observing that 
\be 
\langle \Omega\, ,\, \e^{\ri t H}\Omega \rangle = 
\langle \Omega\, ,\, \e^{\frac{\eta t}{2}(b^{\ast 2} - b^2)}\Omega \rangle \, , 
\label{B1}
\ee   
where 
\be 
[b\, ,\, b^\ast ] = 1\, . 
\label{B2}
\ee 
Setting $\cN = -2 - 4 b^\ast b$ one immediately verifies that 
\be 
[b^{\ast 2}\, , b^2]=\cN\, , \qquad [\cN\, , \, b^{\ast 2}] = -8 b^{\ast 2}\, , \qquad 
[\cN\, , \, b^2] = 8 b^2\, ,
\label{B3}
\ee
showing that $\{b^2,\, b^{\ast 2},\, \cN\}$ constitute a generator basis for a realization of the $SU(1,1)$ Lie 
algebra. Therefore, one can rewrite the $SU(1,1)$ group element $\e^{\alpha (b^{\ast 2} - b^2)}$ as a 
product of exponentials of the algebra generators, i.e. 
\be 
\e^{\alpha (b^{\ast 2} - b^2)} = \e^{f(\alpha)b^{\ast 2}}\,  \e^{g(\alpha)b^2}\,  \e^{h(\alpha)\cN}\, , 
\label{B4}
\ee 
where $f, g, h$ are some functions of the real variable $\alpha$, satisfying 
\be 
f(0) = g(0) = h(0) =0 \, . 
\label{B5}
\ee
We are particularly interested in $h$, because 
\be  
\langle \Omega\, ,\, \e^{\frac{\eta t}{2}(b^{\ast 2} - b^2)}\Omega \rangle 
= \e^{-2h\left (\frac{\eta t}{2}\right )} \, . 
\label{B6}
\ee 
From group theory we know that $f, g, h$ satisfy first order coupled differential equations. 
They are easily obtained by deriving both sides of eq. (\ref{B4}) with respect to $\alpha$ and 
identifying the coefficients of $b^{\ast 2}$, $b^2$ and $\cN$ after shifting to the right the 
remaining exponential part. In this way one gets  
\bea 
\frac{\rd f}{\rd \alpha}  &=& 1-4f^2\, , 
\nonumber \\
\frac{\rd g}{\rd \alpha} &=& 8fg-1\, , 
\nonumber \\
\frac{\rd h}{\rd \alpha} &=& f\, . 
\label{B7}
\eea
The solution of this system with initial condition (\ref{B5}) is 
\be 
f(\alpha ) = \frac{1}{2} \tanh (2\alpha)\, , \qquad 
g(\alpha ) = -\frac{1}{4} \sinh (4\alpha)\, , \qquad 
h(\alpha ) = \frac{1}{4} \ln\left [\cosh (2\alpha)\right ]\, , 
\label{B8}
\ee
which, combined with (\ref{B6}), implies (\ref{vte}). 

In order to derive (\ref{cstate2}) one can use the identity 
\be 
\langle \e^{zb^\ast - {\overline z} b} \Omega\, ,\, \e^{\alpha (b^{\ast 2} - b^2)}\, \e^{zb^\ast - {\overline z} b} \Omega \rangle 
= \e^{|z|^2} \langle \Omega\, ,\, \e^{{\overline z} b}\, \e^{f(\alpha)b^2}\,  
\e^{g(\alpha)b^{\ast 2}}\,  \e^{h(\alpha)\cN}\, \e^{zb^\ast} \Omega \rangle \, , 
\label{B9}
\ee 
which follows from (\ref{B2}), (\ref{B4}) and $b\Omega =0$. The strategy for computing the right hand side of 
(\ref{B9}) is simply to move the exponents involving $\cN$ and $b^2$ and $b$ to the right by using the commutation 
relations among the operators in the exponents. After some algebra one obtains (\ref{cstate2}). 

\sect{BBS contribution and Kirchhoff's rules} 

We show here that 
\be 
\sum_{i=1}^n U_{ij} = \pm 1\quad  \Longrightarrow \quad \sum_{i=1}^n \U_{ij} = 0 \, , 
\label{b1}
\ee 
{}for all $j$ such that $\alpha_j \not= 0, \pm \frac{\pi}{2}$ in (\ref{d2}). Rewriting
\be 
U_d=\U^{-1}\, U\, \U 
\label{b2}
\ee  
in the form 
\be 
\sum_{k=1}^n \U_{ik} (U_d)_{kj} = \sum_{k=1}^n U_{ik}\, \U_{kj} 
\label{b3}
\ee
and using that 
\be 
(U_d)_{kj} = \delta_{kj}\e^{2\ri \alpha_k} 
\label{b4}
\ee 
one gets 
\be 
\U_{ij} \e^{2\ri \alpha_j} = \sum_{k=1}^n U_{ik}\, \U_{kj}\, . 
\label{b5}
\ee 
Summing over $i$ from 1 to $n$ one obtains from (\ref{b5}) 
\be 
\sum_{i=1}^n \U_{ij} \e^{2\ri \alpha_j} = \sum_{i=1}^n \sum_{k=1}^n U_{ik}\, \U_{kj}\, .
\label{b6}
\ee 
Combining the hypothesis in (\ref{b1}) with (\ref{b6}) and using $U^t = U$, one finds 
\be 
\sum_{i=1}^n \U_{ij} \e^{2\ri \alpha_j} = \pm \sum_{k=1}^n \U_{kj}\, , 
\label{b7}
\ee 
or, equivalently 
\be 
(\e^{2\ri \alpha_j} \mp 1)\sum_{i=1}^n \U_{ij} = 0\, , 
\label{b8}
\ee 
which proves the statement (\ref{b1}).

\end{document}